\documentclass[11pt]{article}

\usepackage[final]{acl}
\usepackage{enumitem}
\usepackage{times}
\usepackage{latexsym}
\usepackage{multirow}
\usepackage{amsmath}

\usepackage[T1]{fontenc}

\usepackage[utf8]{inputenc}

\usepackage{microtype}
\usepackage{listings}
\usepackage{xcolor}

\usepackage{inconsolata}

\usepackage{graphicx}

\usepackage{cleveref}
\usepackage{booktabs}
\usepackage{adjustbox}

\title{{\sc REaR}: Retrieve, Expand and Refine for Effective Multitable Retrieval}

\author{
Rishita Agarwal$^{1*}$,
Himanshu Singhal$^{1*}$,
Peter Baile Chen$^{2\dagger}$,
Manan Roy Choudhury$^{3\dagger}$\\
\textbf{Dan Roth}$^{4}$, 
\textbf{Vivek Gupta}$^{3}$
\\[4pt]
$^{1}$ Indian Institute of Technology Guwahati \quad
$^{2}$ Massachusetts Institute of Technology \\ \quad
$^{3}$ Arizona State University \quad
$^{4}$ University of Pennsylvania
\\[2pt]
\texttt{vgupt140@asu.edu}
\\[2pt]
{\small $^{*}$Equal contribution (co-first authors)\quad
$^{\dagger}$Equal contribution (co-second authors)}
}

\begin{document}
\maketitle

\begin{abstract}
Answering natural language queries over relational data often requires retrieving and reasoning over multiple tables, yet most retrievers optimize only for query–table relevance and ignore table–table compatibility. We introduce {\sc REaR} (Retrieve, Expand and Refine), a three-stage, LLM-free framework that separates semantic relevance from structural joinability for efficient, high-fidelity multi-table retrieval. {\sc REaR} (i) retrieves query-aligned tables, (ii) expands these with structurally joinable tables via fast, precomputed column-embedding comparisons, and (iii) refines them by pruning noisy or weakly related candidates. Empirically, {\sc REaR} is retriever-agnostic and consistently improves dense/ sparse retrievers on complex table QA datasets (BIRD, MMQA, and Spider) by improving both multi-table retrieval quality and downstream SQL execution. Despite being LLM-free, it delivers performance competitive with state-of-the-art LLM-augmented retrieval systems (e.g., ARM) while achieving much lower latency and cost. Ablations confirm complementary gains from expansion and refinement, underscoring {\sc REaR} as a practical, scalable building block for table-based downstream tasks (e.g., Text-to-SQL).




\end{abstract}

\section{Introduction}
\label{sec:intro}



\begin{figure}[t]
    \centering
    \includegraphics[width=0.97\linewidth]{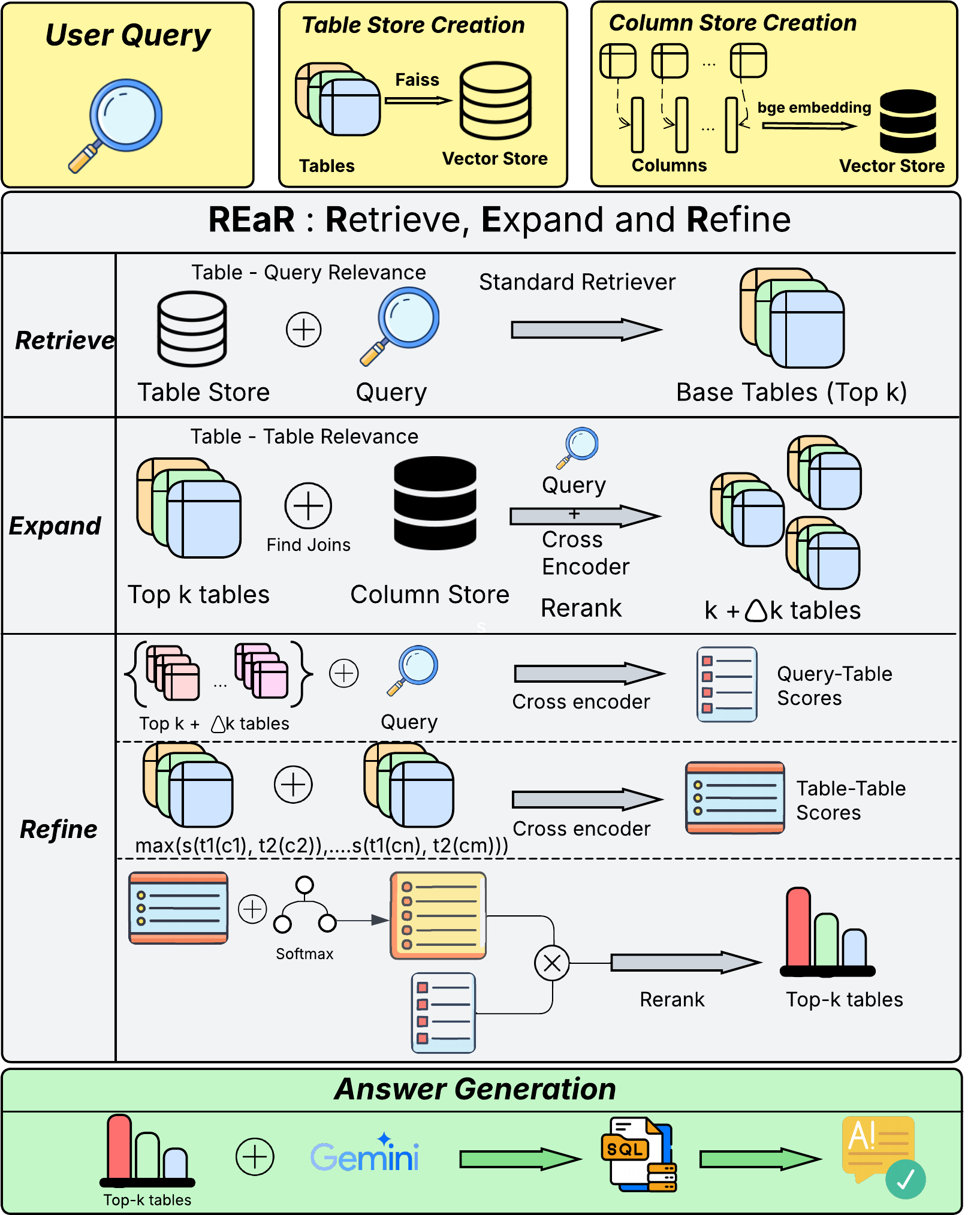}
    \vspace{-0.35em}
    \caption{Overview of the {\sc REaR} framework. (1) \textbf{Retrieve}: Select top-\textit{k} relevant tables. (2) \textbf{Expand:} Augment with joinable tables using FAISS column search and cross-encoder reranking (\textit{k} $\to$ \textit{k'}$+\Delta k'$ candidates). (3)\textbf{ Refine:} Score candidates by query relevance and table joinability, rerank to final top-\textit{k}. Top boxes: offline preprocessing}
    \label{fig:overview}
    \vspace{-1.75em}
\end{figure}

Retrieval in current Text-to-SQL pipelines is fundamentally myopic: it optimizes for query–table relevance while ignoring table–table compatibility. This mismatch creates a brittle handoff, retrievers surface tables that look topically aligned, but lack the join paths, key constraints, or schema alignment needed for integration. Downstream, the semantic parser inherits an incoherent candidate set, forcing it to guess joins, drop constraints, or hallucinate links, errors that propagate into invalid or incomplete SQL. In short, relevance-only retrieval starves the parser of relational context; without explicit reasoning over joinability and connectivity, even a “relevant” set of tables cannot be composed into a correct query. 
The remedy is retrieval that jointly scores (i) topical fit and (ii) relational fit, so Text-to-SQL starts from a schema-aware, join-ready subset rather than a bag of loosely related tables.

Prior LLM-based efforts span schema linking (E-SQL \cite{caferouglu2024sql}, RSL-SQL \cite{cao2024rsl}), column-level joinability (DeepJoin \cite{dong2023deepjoin}), and multi-hop retrieval (MURRE \cite{zhang-etal-2025-murre}), yet they often falter when relationships are implicit or distant. Despite these advances, gaps remain: many techniques require heavy, dataset-specific training that weakens generalization to unseen databases; multi-hop strategies can be computationally expensive; and overall, they still trade off recall (capturing all necessary tables) against precision (excluding distractors), limiting real-world utility. More recently, training-free retrieval frameworks such as CRAFT~\cite{singh2025crafttrainingfreecascadedretrieval} have explored cascaded retrieval for tabular question answering, emphasizing modularity and efficiency without fine-tuning. However, these methods primarily focus on cascade design rather than explicitly modeling table–table joinability. Meanwhile, state-of-the-art reasoning-augmented retrieval methods (e.g., ARM~\citep{chen-etal-2025-retrieve}) rely on costly LLM calls, trading efficiency for stronger performance. This raises a critical question: \emph{Can we design a \textbf{simple and efficient retriever} that scores both \textbf{query–table relevance and table–table} compatibility?}


To address this question, we introduce {\sc REaR} (\textbf{R}etrieve, \textbf{E}xpand \textbf{a}nd \textbf{R}efine), a three-stage framework that cleanly separates query–table from table–table reasoning to deliver efficient, high-fidelity multi-table retrieval without any LLM calls.
As illustrated in \Cref{fig:overview}, {\sc REaR} (i) retrieves tables that are semantically aligned with the query, (ii) expands this set by adding tables that are structurally joinable with the retrieved ones, and (iii) refines the pool by pruning noisy or redundant items. This decomposition jointly captures what information is needed (query–table relevance) and how it connects (table–table compatibility), addressing core limitations of prior methods. 

Concretely, the \textit{Expansion} stage boosts recall by scanning the database for tables that are semantically joinable with this set via precomputed column-embedding comparisons that are far more efficient than LLM-based checks to uncover latent relationships required for complex joins. Furthermore, the \textit{Refinement} stage removes irrelevant or weakly related tables, producing a final set that is both precise and complete. Overall, {\sc REaR} forms a lightweight pipeline that unites semantic and structural reasoning for efficient, LLM-free multi-table retrieval. Across multiple complex table QA datasets, {\sc REaR} reliably improves standard retrievers, reaching performance close to state-of-the-art LLM-based retrieval systems (e.g., JAR~\citep{chen-etal-2024-table} and ARM) while remaining compute- and cost-efficient. By better selecting and refining tables, {\sc REaR} also boosts downstream SQL execution performance.

\noindent Our contributions are as follows:

\begin{itemize}[leftmargin=*,itemsep=0.1em]

\item We introduce \textsc{REaR}, which separates query–table relevance from table–table joinability, using precomputed column embeddings to identify join-ready tables before effective precision filtering.

\item \textsc{REaR} eliminates LLM calls from retrieval, enhancing retrieval performance with substantially lower latency and cost, making it practical for large-scale real-world usage.

\item \textsc{REaR} acts as a plug-and-play layer over standard retrievers, boosting recall (through expansion) and precision (through refinement) without modifying the base retriever.

\item Extensive experiments on MMQA, BIRD, and Spider show that \textsc{REaR} also improves downstream SQL execution accuracy, with ablations validating the impact of each stage.

\end{itemize}



\section{Methodology}

Inspired by~\cite{chen-etal-2024-table, chen-etal-2025-retrieve, dong2023deepjoin}, an effective multi-table retriever should jointly optimize query–table relevance and table–table joinability. The former ensures that retrieved tables contain information germane to the user query, whereas the latter guarantees that the tables can be connected so that their evidence can be composed to answer the query.
As discussed in \Cref{sec:intro}, existing approaches frequently employ LLMs during retrieval, which is often impractical due to latency and cost. To mitigate these limitations, we propose a top-$k$ multi-table retrieval pipeline that avoids online LLM usage by explicitly models both objectives using offline embeddings of tables, queries, and columns.

Given a query $q$, {\sc REaR} comprises three stages: retrieval, expansion, and refinement.
(1) \textbf{Retrieval:} a standard dense, sparse, or hybrid retriever selects $k'$ base tables that are semantically relevant to $q$, establishing query–table relevance.
(2) \textbf{Expansion:} $\Delta k'$ additional tables that are joinable with the base set are introduced, promoting table–table joinability.
(3) \textbf{Refinement:} the $(k' + \Delta k')$ candidates are reduced to $k$ by jointly scoring relevance and joinability, and the resulting set is returned as the final top-$k$ tables. Detailed specifications of these stages are provided in \Cref{sec:retrieve}, \Cref{sec:expand}, and \Cref{sec:refine}.




\subsection{Retrieval}
\label{sec:retrieve}


For a user query $q$, the retrieval stage first identifies a set of \textit{base tables} $\mathcal{T}_{\text{base}} \subset \mathcal{T}$ that are semantically relevant to $q$ to ensure query-table relevance, where $\mathcal{T}$ denotes the full table corpus. These base tables provide the foundation for subsequent stages, which further account for table-table joinability.

Specifically, we compute a relevance score $s(q, t_i)$ for each table $t_i \in \mathcal{T}$, rank all tables by their scores in descending order, and select top-$k'$ tables: $\mathcal{T}_{\text{base}} = \{t_{(1)}, t_{(2)}, \ldots, t_{(k')}\}$, where $t_{(i)}$ denotes the $i$-th highest-scoring table. The relevance scoring function $s(q, t_i)$ varies by retriever type:

\begin{enumerate}[leftmargin=*,noitemsep]
    \item {\textbf{Sparse retrievers}} compute relevance using TF-IDF scores based on term overlap between query and table.
    \item {\textbf{Dense retrievers}} compute relevance as cosine similarity between learned query and table embeddings.
    \item \textbf{Hybrid retrievers} combine both approaches:
\vspace{-0.35em}
\begin{equation*}
\noindent
s_{\text{hybrid}}(q, t_i) =
\begin{aligned}[t]
    &\alpha \cdot s_{\text{sparse}}(q, t_i) \\
    &\quad + (1-\alpha)\cdot s_{\text{dense}}(q, t_i)
\end{aligned}
\end{equation*}
where $\alpha \in [0, 1]$ is a weighting hyperparameter.
\end{enumerate}
Further details on query-table relevance computation are provided in \Cref{app:qt relevance}.


\subsection{Expansion}
\label{sec:expand}

Although the base tables $\mathcal{T}_{\text{base}}$ described in \Cref{sec:retrieve} are semantically relevant to the user query $q$, they may not be mutually joinable, which limits their ability to compose information to address $q$. To address this, we introduce an expansion stage that explicitly adds tables joinable with the base set, ensuring table-table joinability and thereby increasing the likelihood of answering $q$.
Concretely, the expansion stage consists of two steps: detection and reranking. We first identify tables from the corpus $\mathcal{T}$ that are joinable with the base tables, and then rerank these candidates by their query-table relevance, yielding tables that are both joinable and semantically relevant to the query.

\paragraph{Detection.}
Inspired by DeepJoin~\cite{dong2023deepjoin}, we identify joinable tables by measuring the similarity between column embeddings from pairs of tables. Formally, let $\mathcal{C}_{1}(t)$ denote the set of columns in table $t$. For two tables $t_i, t_j \in \mathcal{T}$, we define them as joinable if:
\vspace{-0.75em}
\begin{equation*}
\small
\text{joinable}(t_i, t_j) = \max_{c_i \in \mathcal{C}_{1}(t_i), c_j \in \mathcal{C}_{1}(t_j)} \text{sim}(c_i, c_j) \geq \tau
\vspace{-0.75em}
\end{equation*}
where $\text{sim}(c_i, c_j)$ is the cosine similarity between column embeddings and $\tau$ is a threshold (set to $0.7$ in our experiments). Unlike DeepJoin, which relies on dataset-specific embedding models, we use general-purpose pre-trained embeddings to ensure cross-domain generalizability.

To construct a column embedding $\mathbf{e}_c$ for column $c$, we first serialize it by concatenating its name and values into a text representation $\text{serialize}(c)$, which is then encoded using a dense embedding model $f_{\text{enc}}$:
\vspace{-0.5em}
\begin{equation*}
\small
\mathbf{e}_c = f_{\text{enc}}(\text{serialize}(c))
\vspace{-0.5em}
\end{equation*}

Since exhaustively comparing all column pairs requires $\mathcal{O}(|\mathcal{T}|^2 \cdot |\mathcal{C}_{\text{avg}}|^2)$ operations (where $|\mathcal{C}_{\text{avg}}|$ is the average number of columns per table), we leverage approximate nearest neighbor search algorithms~\citep{douze2024faiss}, enabling sublinear-time retrieval of candidate column matches.

Finally, for each base table $t_b \in \mathcal{T}_{\text{base}}$ obtained in \Cref{sec:retrieve}, we enumerate its joinable tables to form the candidate joinable set:
\vspace{-0.5em}
\begin{equation*}
\small
\mathcal{T}_c = \bigcup_{t_b \in \mathcal{T}_{\text{base}}} \{t \in \mathcal{T} \mid \text{joinable}(t_b, t)\}
\vspace{-0.5em}
\end{equation*}

\paragraph{Reranking.}
To filter tables that are semantically distant from the user query, we rerank the candidate joinable tables $\mathcal{T}_c$ using a cross-encoder reranker $r(q, t)$ that computes joint query-table representations. We rank all tables in $\mathcal{T}_c$ by their reranker scores and select the top-$(\Delta k')$:
\vspace{-0.5em}
\begin{equation*}
\small
\mathcal{T}_{\text{join}} = \{t_{(1)}, t_{(2)}, \ldots, t_{(\Delta k')}\}
\vspace{-0.5em}
\end{equation*}
where $t_{(i)}$ denotes the $i$-th highest-scoring table under $r(q, \cdot)$. These filtered joinable tables are combined with the base tables to form the expanded set:
\vspace{-0.5em}
\begin{equation*}
\small
\mathcal{T}_{\text{expanded}} = \mathcal{T}_{\text{base}} \cup \mathcal{T}_{\text{join}}
\vspace{-0.5em}
\end{equation*}

\subsection{Refinement}
\label{sec:refine}

Finally, this stage is designed to restore precision. While expansion adds potentially useful but \textit{noisy} tables, refinement filters and reprioritizes these candidates to produce a smaller, high-quality set of tables that are both semantically relevant to the query and structurally coherent with each other.
We refine the expanded set $\mathcal{T}_{\text{expanded}}$ (described in \Cref{sec:expand})
by selecting the final subset of tables to return, \textit{jointly} considering query–table relevance and table–table joinability.


For each table $T_i$ in the expanded set, we compute a score $S(T_i)$, which is then used to rank all tables. The top-$k$ tables with highest scores $S(T_i)$ are returned, where $S(T_i)$ combines query-table relevance and table-table joinability:

\vspace{-0.5em}
\begin{equation*}
\small
S(T_i) = C_{2}(q, T_i) \cdot A(T_i)
\vspace{-0.25em}
\end{equation*}
where $C_{2}(q, T_i)$ and $A(T_i)$ denote the query-table relevance and table-table joinability, respectively, and $q$ is the user query.


\paragraph{Query-table relevance.}
We first compute the similarity between the query $q$ and $T_i$ using a cross-encoder model (detailed in Section~\ref{sec:exp-setup}), yielding the score $C_{2}(q, T_i)$. We employ a cross-encoder rather than embedding-based similarity for two key reasons. First, cross-encoders enable fine-grained token-level interactions between the query and table schema through full self-attention, allowing the model to identify subtle semantic relationships that bi-encoders, which encode inputs independently, cannot capture. Second, at this stage, we operate on a small candidate set, making the computational overhead of cross-encoders tractable while prioritizing scoring accuracy over inference speed. This mirrors standard practice: bi-encoders for retrieval, cross-encoders for reranking.
 
\paragraph{Table-table joinability.}
We then assess the relational coherence of $T_i$ with respect to other tables in the retrieved set.
\vspace{-0.25em}
\begin{equation*}
\small
A(T_i) = \max( \text{softmax}(C_{2}(T_i, T_j)) \cdot C_{2}(T_i, T_j))
\vspace{-0.25em}
\end{equation*}

For each table pair $(T_i, T_j)$ where $T_j$ is in the neighborhood of $T_i$, we compute column-level similarities using the same cross-encoder. Specifically, for all column pairs between $T_i$ and $T_j$, we retain the maximum similarity score as the table-pair score $C_{2}(T_i, T_j)$. This design leverages the observation that joinable columns sharing foreign key relationships exhibit high semantic similarity.


We compute the attention score $A(T_i)$ by applying softmax normalization over all table-table similarity scores $C_{2}(T_i, T_j)$, selecting the maximum normalized score, and scaling it by the corresponding similarity value. This max operation suppresses contributions from weakly-related tables.
\section{Experiments}
\label{sec:exp}

Our goal is to evaluate the effectiveness of the proposed multi-table retrieval pipeline, {\sc REaR}, which introduces a novel retrieve–expand–refine process. Compared to \textit{standard retrievers} that lack the expansion and refinement stages, we aim to understand the added benefits of these components in our pipeline. Unlike \textit{LLM-based retrieval} methods, our approach avoids online LLM calls entirely.

For evaluation, we use complex open-domain Text-to-SQL datasets as they require reasoning over multiple tables to produce correct answers. These datasets allow us to assess performance along two axes: (1) whether gold tables are retrieved, capturing retrieval performance, and (2) whether the correct final answer is generated, capturing end-to-end performance.



\subsection{Experimental Setup}
\label{sec:exp-setup}

\paragraph{Datasets.}
We evaluate all methods on complex text-to-SQL datasets that require reasoning over multiple tables: MMQA~\cite{wu2025mmqa}, Spider~\cite{yu-etal-2018-spider}, and Bird~\cite{li2023can}. Since most databases in Spider and Bird contain at most 10 tables, retrieval in these cases is trivial: for example, retrieving all 10 tables from a 10-table database guarantees 100\% recall. To create a more challenging evaluation setting, we merge all databases into a single corpus (statistics in \Cref{tab:dataset_stats}).



\begin{table}[t]
\setlength{\aboverulesep}{0pt}
\setlength{\belowrulesep}{0pt}
\centering
\small
\caption{Dataset statistics and characteristics}
\vspace{-0.75em}
\label{tab:dataset_stats}
\setlength{\tabcolsep}{3pt}
\begin{tabular}{l|rrrr}
\hline
& \bf BIRD & \bf SPIDER & \bf MMQA \\
\cmidrule(lr){2-4}
\bf Avg. Rows per Table & 52436.46 & 6742.46 & 1732.24 \\
\bf Avg. Columns per Table & 10.64 & 5.51 & 5.77 \\
\bf Avg. Tables per Query & 1.96 & 1.91 & 2.22 \\
\bf Total Databases & 11 & 20 & -- \\
\bf Total Tables & 75 & 139 & 695 \\
\bf Total Queries & 1534 & 1034 & 3312 \\
\hline
\end{tabular}
\vspace{-1.0em}
\end{table}


\begin{table*}[!htb]
\setlength{\aboverulesep}{0pt}
\setlength{\belowrulesep}{0pt}
\setlength{\tabcolsep}{3.3pt}
\small
\centering
\caption{Retrieval performance of standard retrievers and our {\sc REaR} pipeline (\textit{``Retrieval''}, \textit{``+Expansion''}, \textit{``+Refining''}). The expansion stage of our method expands the set of candidate tables to 8, and the refinement stage reduces the set of tables to 5.
\textbf{Bold} and \underline{underline} numbers denote best and second-best, respectively.}
\vspace{-0.75em}
\begin{tabular}{l | cc c | cc c | cc c}
\hline
 & \multicolumn{3}{c}{\bf BIRD} & \multicolumn{3}{c}{\bf MMQA} & \multicolumn{3}{c}{\bf Spider} \\
\cmidrule(lr){2-4} \cmidrule(lr){5-7} \cmidrule(lr){8-10}
 & \bf Recall & \bf Precision & \bf Full Recall & \bf Recall & \bf Precision & \bf Full Recall & \bf Recall & \bf Precision & \bf Full Recall \\
\hline
& \multicolumn{9}{c}{\textbf{Dense Retrievers}} \\  
\bf  BGE (Top 8) & \underline{96.11} & 23.88 & \underline{91.53} & 79.29 & 21.94 & 60.18 & \underline{97.92} & 18.54 & \underline{96.62} \\
Retrieval (Top 5) & 90.70 & \underline{35.64} & 81.75 & 72.03 & \underline{31.78} & 48.53 & 96.69 & \underline{29.20} & 94.19 \\
~~~~+Expansion              & \textbf{96.51} & 24.66 & \textbf{93.16} & \textbf{87.40} & 24.31 & \textbf{74.71} & \textbf{98.84} & 18.76 & \textbf{98.55} \\
~~~~+Refining ({\sc REaR})          & 93.32 & \textbf{35.90} & 86.42 & \underline{82.78} & \textbf{36.70} & \underline{65.83} & 96.69 &\textbf{ 29.26} & 95.94 \\
\hline
\bf  UAE (Top 8)                  & \underline{94.72} & 23.92 & \underline{89.37} & 81.28 & 22.49 & 64.29 & \underline{99.10} & 18.77 & \underline{98.45} \\
Retrieval (Top 5)                 & 90.68 & \underline{36.09} & 82.79 & 73.92 & \underline{32.61} & 52.27 & 98.06 & \textbf{29.75} & 96.61 \\
~~~~+Expansion             & \textbf{96.81} & 25.26 & \textbf{93.42} & \textbf{87.91} & 24.44 & \textbf{75.64} & \textbf{99.32} & 18.88 & \textbf{99.13} \\
~~~~+Refining ({\sc REaR})                  & 93.96 & \textbf{36.33} & 87.29 & \underline{83.19} & \textbf{36.86} & \underline{66.61} & 98.50 & \underline{29.61} &{97.68} \\
\hline
\bf  GTE (Top 8)                   & 92.54 & 22.82 & \underline{85.40} & 73.25 & 20.17 & 53.34 & \underline{98.03} & 18.59 & \underline{97.00} \\
Retrieval (Top 5)                   & 87.46 & \underline{34.05} & 76.60 & 64.95 & \underline{28.52} & 41.93 & 94.25 & \underline{28.22} & 89.85 \\
~~~~+Expansion              & \textbf{96.17} & 24.24 & \textbf{92.13} & \textbf{85.72} & 23.80 & \textbf{72.53} & \textbf{98.79} & 18.75 &\textbf{ 98.45} \\
~~~~+Refining ({\sc REaR})               & \underline{92.85} & \textbf{35.70} & 84.95 & \underline{81.65} & \textbf{36.17} & \underline{64.32} & 97.10 & \textbf{29.28} & 96.03 \\
\hline
\bf e5 (Top 8)                  &\underline{95.17} & 22.90 & \underline{89.57} & 77.75 & 21.68 & 58.59 & \underline{97.59} & 18.38 & \underline{96.22} \\
Retrieval (Top 5)                  & 90.49 & \underline{34.41} & 81.49 & 69.96 & \underline{31.09} & 45.94 & 94.96 & \underline{28.45} & 91.50 \\
~~~~+Expansion             & \textbf{95.38} & 23.25 & \textbf{90.66} & \textbf{87.40} & 24.31 & \textbf{74.74} & \textbf{98.59} & 18.68 & \textbf{98.16} \\
~~~~+Refining ({\sc REaR})                 & 92.18 & \textbf{35.40} & 83.82 & \underline{83.57} & \textbf{37.08} & \underline{67.37} & 96.94 & \textbf{29.23} & 95.84 \\
\hline
& \multicolumn{9}{c}{\textbf{Sparse Retrievers}} \\  
\bf  BM25 (Top 8)                 & 88.28 & 21.41 & 77.25 & 79.32 & 22.09 & 58.90 & 95.08 & 17.84 & 91.10 \\
Retrieval (Top 5)                 & 81.10 & \underline{31.21} & 65.65 & 73.16 & \underline{32.48} & 48.42 & 91.02 & \underline{27.16} & 84.82 \\
~~~~+Expansion             & \textbf{93.71} & 23.08 & \textbf{88.40} & \textbf{87.62} & 24.40 & \textbf{75.03} & \textbf{98.55} & 18.70 & \textbf{98.07} \\
~~~~+Refining ({\sc REaR})               & \underline{91.04} & \textbf{34.86} & \underline{82.72} & \underline{83.69} & \textbf{37.16} & \underline{67.43} & \underline{96.76} & \textbf{29.21} & \underline{95.74} \\
\hline
\bf SPLADE (Top 8)               &  \underline{96.02}     &  23.83     & \underline{91.4}      &   82.18    &  22.82     &  64.26     &  91.04     &  17.14     &  86.84     \\
Retrieval (Top 5)               &  90.97      &  \underline{35.69}       & 81.75  &  75.91     & \underline{33.57}      & 53.9     &  81.14    &   \underline{24.31}      &  71.76        \\ 
~~~~+Expansion          &       \textbf{97.67}     & 24.07    &     \textbf{94.58}   &   \textbf{88.99}    &   24.76    &  \textbf{77.66}     &  \textbf{99.17}     &   18.65   & \textbf{98.83}      \\
~~~~+Refining ({\sc REaR})               & 94.15     &   \textbf{36.08}    &   87.16    &    \underline{83.86}   &  \textbf{37.16}     &  \underline{67.79}     &   \underline{97.22}    &   \textbf{29.34}    &  \underline{96.32}     \\

\hline
& \multicolumn{9}{c}{\textbf{Hybrid Retrievers}} \\  
\bf SPLADE$_{\text{Hybrid}}$ (Top 8)               &  \underline{96.08}     & 23.83    &  \underline{91.05}     & \underline{84.82}       & 23.5      & \underline{68.97}      &  95.48     &  24.97     & 93.35      \\
Retrieval (Top 5)               & 92.49      & \underline{35.96} & 84.06     &  78.4     &  \underline{34.59}     & 57.13      &  85.46     & \textbf{29.46}      &  75.04     \\ 
~~~~+Expansion              &  \textbf{97.27}     &  23.88     & \textbf{93.81}      & \textbf{ 89.19}     &  24.79     &  \textbf{77.69}     & \textbf{96.74}      & 18.24      & \textbf{95.55}      \\
~~~~+Refining ({\sc REaR})                 &   93.88    &  \textbf{36.43}     &  86.64     &  83.71     &  \textbf{37.11 }    & 67.2      &  \underline{96.47}     &  \underline{29.07}     &  \underline{94.97}     \\

\hline
\end{tabular}
\label{tab:standard-retrieval}
\vspace{-1.5em}
\end{table*}
\paragraph{Baselines.}
We compare our method with multiple standard and LLM-based retrieval methods.

- For \textbf{standard retrievers}, we consider three methods: sparse retrievers, dense retrievers, and hybrid methods combining both.
For sparse retrievers, we employ the BM25 implementation from \cite{10.1561/1500000019} and Splade~\cite{10.1145/3477495.3531833}. For dense retrievers, we select lightweight models that rank highly on the MTEB leaderboard, including UAE-Large-V1~\cite{li-li-2024-aoe}, GTE-large~\cite{unknown}, e5-mistral~\cite{wang2023improving}, and BGE-large-en~\cite{bge_embedding}. For hybrid retrievers, we used Splade retriever's hybrid version for a vast coverage of different types of retrievers.


- For \textbf{LLM-based retrieval} methods, we evaluate two approaches: JAR~\citep{chen-etal-2024-table} and ARM~\cite{chen-etal-2025-retrieve}.
These methods dynamically determine the number of tables for SQL generation rather than fixing it a priori.

\paragraph{Evaluation Metrics.}
We assess the performance of different methods along two dimensions: retrieval performance and end-to-end performance.

- For \textbf{retrieval performance}, we report standard precision and recall, along with a stricter metric we call full recall, by comparing the retrieved tables and the gold tables. Full Recall is a binary measure indicating whether all gold tables appear in the retrieved set, a necessary condition for generating the correct final answer.

- For \textbf{end-to-end performance}, we use execution accuracy, which evaluates query correctness by comparing the execution results of SQL statements generated using the retrieved tables against those of the gold SQL statements. The accuracy is 1 if the results are the same and 0 otherwise.

\paragraph{Implementation details.}
We generate table descriptions to augment the original table schemas for all methods, inspired by ~\cite{chen2025enrichindex}.



For column encoding in the expansion stage, we evaluated several state-of-the-art embedding models from the MTEB leaderboard, including bge-large-en, Qwen-Embedding~\cite{zhang2025qwen3}, and Gemini-Embeddings~\cite{lee2025gemini}. Based on our experiments, we adopted \textbf{bge-large-en}, which achieves performance comparable to larger models while maintaining computational efficiency. (More details in \cref{sec:appendix})

We fixed the number of joinable tables to 3 for all main experiments. We do this since most queries in our target datasets require 3 or less than 3 tables to answer the questions and to maintain the computational efficiency of our pipeline. We employ Jina-reranker as the reranker, a state-of-the-art and computationally efficient reranking model. In the refinement stage of our retrieval pipeline, we used Jina-reranker as the cross-encoder. For SQL generation, we employ both small and large language models: Gemini-2.0-Flash, Llama-3.2-3B, and Gemma-3-4B.






\subsection{Results and analysis}
\label{sec:exp-result}
\begin{table*}[!htb]
\small
\setlength{\aboverulesep}{0pt}
\setlength{\belowrulesep}{0pt}
\setlength{\tabcolsep}{7pt}
\centering
\caption{SQL execution accuracy at each stage of the {\sc REaR} pipeline (\textit{``Retrieval''}, \textit{``+Expansion''}, \textit{``+Refining''}) compared to baseline retrieval methods. UAE-V1-Large is used as the dense retriever for optimal performance, with Splade-Sparse and Splade-Hybrid as the sparse and hybrid retrievers, respectively. \textbf{Bold} and \underline{underline} numbers denote best and second-best, respectively, excluding Oracle.}
\vspace{-0.75em}
\label{tab:standard-ete}
\begin{tabular}{l| ccc | ccc | ccc}
\hline
\textbf{} & \multicolumn{3}{c}{\textbf{BIRD}} & \multicolumn{3}{c}{\textbf{SPIDER}} & \multicolumn{3}{c}{\textbf{MMQA}} \\ 
\cmidrule(lr){2-4} \cmidrule(lr){5-7} \cmidrule(lr){8-10}
\textbf{} &\bf Dense &\bf Sparse &\bf Hybrid &\bf Dense &\bf Sparse &\bf Hybrid &\bf Dense &\bf Sparse &\bf Hybrid \\ 
\hline 
& \multicolumn{9}{c}{\textbf{Gemini 2.0 Flash}} \\
Oracle Retrieval &  & 50.39 &  &  &  76.98&  &  & 53.9 &  \\ 
\cmidrule(lr){2-4} \cmidrule(lr){5-7} \cmidrule(lr){8-10}
Retrieval (Top 8) & 38.33 & 41 & \underline{41.91} & \underline{69.54} & 64.02 & 70.79 & 30.67 & 33.19 & 35.16 \\ 
~~~Oracle Prune & 46.21 & 45.83 & 46.98 & 76.20 & 68.76 & 74.75 & 43.83 & 37.94 & 42.37 \\ \cmidrule(lr){2-4} \cmidrule(lr){5-7} \cmidrule(lr){8-10}
Retrieval (Top 5) & 37.87 & 37.03 & 39.84 & 69.44 & 57.54 & 46.9 & 26.73 & 30.82 & 32.96 \\ 
~~+Expansion & \underline{42.24} & \underline{41.13} & \textbf{42.91} & 69.34 & \underline{70.79} & \underline{70.9} & \underline{37.45} & \underline{37.84} & \underline{37.88} \\  
~~~~~+Refining ({\sc REaR}) & \textbf{43.09} & \textbf{41.54} & 40.76 &\textbf{ 69.93} & \textbf{70.8} & \textbf{71.8} & \textbf{38.87} & \textbf{37.87} & \textbf{38.15} \\
~~~~~~Oracle Prune &  47.96 & 49.15 &  48.72 & 76.50 & 76.46 & 74.95 & 43.92 & 43.74 & 47.08\\ 
\hline
& \multicolumn{9}{c}{\textbf{Llama-3.2-3b}} \\
Oracle Retrieval &  & 15.71 &  &  & 40.03 &  &  & 30.12 &  \\ 
\cmidrule(lr){2-4} \cmidrule(lr){5-7} \cmidrule(lr){8-10}
Retrieval (Top 8) & 9.12 & 9.58 & 10.12 & 17.41 & 16.24 & 17.98 & 12.85 & 12.24 & \underline{13.94} \\ 
~~~Oracle Prune & 14.46 & 14.64 & 15.07 & 39.55 & 35.1 & 39.36 & 22.3 & 22.42 & 24.02 \\ 
\cmidrule(lr){2-4} \cmidrule(lr){5-7} \cmidrule(lr){8-10}
Retrieval (Top 5) & \underline{9.32} & \underline{11.15} & \underline{10.58}  & \underline{18.96} & 17.69 & 11.41 & \underline{13.28} & \underline{14.15} & 13.85 \\ 
~~+Expansion & 9.13 & 9.45 & 10.05  & 18.57 & \underline{19.24} & \underline{18.08} & 13.09 & 13.4 &  13.91\\ 
~~~~~~+Refining ({\sc REaR}) & \textbf{11.55} & \textbf{11.91} & \textbf{11.38} & \textbf{19.34} & \textbf{19.53} & \textbf{19.05} & \textbf{14.39} & \textbf{14.51} & \textbf{14.12} \\ 
~~~~~~~Oracle Prune & 14.69 & 15.23 & 15.17 & 24.31 & 39.36 & 39.16 & 25.62 & 25.56 & 26.01 \\ 
\hline
& \multicolumn{9}{c}{\textbf{Gemma-3-4b}} \\
Oracle Retrieval &  & 26.27 &  &  & 47.38  & &  & 21.24 &  \\ 
\cmidrule(lr){2-4} \cmidrule(lr){5-7} \cmidrule(lr){8-10}
Retrieval (Top 8) & 13.89 & 12.91 & 14.31 & 17.4 & 17.31 & 15.76 & 8.66 & 8.54 & 8.84 \\ 
~~~Oracle Prune & 24.31 & 24.69 & 25.49 & 47 & 42.06 & 47.29 & 16.4 & 16.39 & 17.11 \\ 
\cmidrule(lr){2-4} \cmidrule(lr){5-7} \cmidrule(lr){8-10}
Retrieval (Top 5) & \underline{17.6} & \underline{16.62} & \underline{17.24}  & \underline{21.76} & \underline{19.92} & \underline{20.11} & \underline{9.8} & \underline{8.99} & \underline{10.26} \\ 
~~+Expansion & 15.67 &13.82  & 17.18 & 16.83 & 18.66 & 17.5 & 8.6 & 8.93 & 9.08 \\ 
~~~~~~+Refining ({\sc REaR}) & \textbf{17.97} & \textbf{17.21} & \textbf{17.64} & \textbf{22.02} & \textbf{24.68} & \textbf{21.4} & \textbf{11.4} & \textbf{10.14} &  \textbf{10.59}\\ 
~~~~~~~Oracle Prune & 24.67 & 25.62 & 25.9 & 47.3 & 46.7 & 47.19 & 18.5 & 18.89 & 18.38 \\ 
\hline
\end{tabular}
\vspace{-1.5em}
\end{table*}


\paragraph{How do the \textit{Expansion} and \textit{Refinement} stages contribute to improvements in retrieval performance over standard top-\textit{k} baselines?}

To isolate the contributions of our pipeline's core components, we first analyze their impact on retrieval metrics. As shown in Table \ref{tab:standard-retrieval}, both stages provide substantial benefits over a standard \textit{``Retrieval (Top 8)''} baseline. Focusing on the BIRD dataset with the UAE dense retriever, the \textbf{Expansion} stage significantly enhances the chance of retrieving the complete set of gold-standard tables, boosting the ``Full Recall'' metric from 82.79\% to 93.42\%, an absolute improvement of 10.63 points. While this stage axiomatically lowers precision by increasing the candidate set size, the subsequent \textbf{Refinement} stage effectively re-optimizes the precision-recall balance. The complete \textit{``+Refining''} pipeline achieves a Full Recall of 87.29\%, which remains nearly 5 points higher than the baseline, while concurrently maintaining precision, i.e., from 36.09\% to 36.33\%. This demonstrates that the refinement mechanism removes noisy expansion candidates while retaining semantically vital tables


\paragraph{Within the Expansion stage, is a join-aware expansion strategy more effective than simply increasing the number of initial retrieved tables?}

A central hypothesis of our work is that a join-aware structural expansion is superior to a naive \textit{k}-augmentation strategy. The data in Table \ref{tab:standard-retrieval} validates this assertion. Employing the BGE retriever on the BIRD dataset, a standard \textit{``Retrieval (Top 8)''} approach yields a Full Recall of 91.53\%. In contrast, our method, which initiates with 5 tables and intelligently expands the set to 8 (\textit{``+Expansion''}), achieves a superior Full Recall of 93.16\%. This pattern across retrievers and datasets shows that modeling inter-table joinability is more effective for recall than increasing the retrieval window.

\begin{table*}
\centering
\small
\caption{REaR against LLM-based methods on retrieval quality and SQL execution accuracy on different datasets across multiple LLMs. "-Desc." represents performance without offline computed LLMs table descriptions. MURRE SQL accuracy is on gpt-3.5-turbo. \textbf{Bold} and \underline{underline} numbers denote best and second-best, respectively; "-" denotes unavailable results.}
\vspace{-0.75em}
\small
\setlength{\aboverulesep}{0pt}
\setlength{\belowrulesep}{0pt}
\setlength{\tabcolsep}{2.35pt}
\begin{tabular}{l|cccccc|cccccc}
\hline 
    \textbf{} & \multicolumn{6}{c|}{\bf BIRD} & \multicolumn{6}{c}{\bf Spider} \\ 
    \textbf{} & \multicolumn{3}{c}{\bf Retrieval} & \multicolumn{3}{c|}{\bf SQL Execution Accuracy} & \multicolumn{3}{c}{\bf Retrieval} & \multicolumn{3}{c}{\bf SQL Execution Accuracy} \\ 
    \bf    & \multicolumn{1}{c}{\bf Precision} & \multicolumn{1}{c}{\bf Recall} & \multicolumn{1}{c|}{\bf Recall$_{\text{Full}}$} & \multicolumn{1}{c}{\bf LLAMA} & \multicolumn{1}{c}{\bf GPT} & \multicolumn{1}{c|}{\bf Gemini} & \bf Precision & \multicolumn{1}{c}{\bf Recall} & \multicolumn{1}{c|}{\bf Recall$_{\text{Full}}$} &  \multicolumn{1}{c}{\bf LLAMA} & \multicolumn{1}{c}{\bf GPT} & \multicolumn{1}{c}{\bf Gemini} \\ \hline
    ARM       & \textbf{42.7} & \underline{96.5} & \multicolumn{1}{c|}{\underline{92.7}} &  20.6 & \textbf{32.4} & 30.4 & \multicolumn{1}{l}{-} & - & \multicolumn{1}{c|}{-} & - & - & - \\
    JAR       & \underline{40.3} & 89.9& \multicolumn{1}{c|}{77.9} & 20.07 & 25.2 & 29.1 & \multicolumn{1}{l}{\textbf{41.9}} & \underline{97.8} & \multicolumn{1}{c|}{\underline{96.23}} & \underline{56.8} & \underline{70.0} & \textbf{75.7} \\ 

     ReAcT      & 15.0 & \textbf{96.7} & \multicolumn{1}{c|}{\textbf{93.5}} & 4.7 & 25.4  & - & \multicolumn{1}{l}{-} & - & \multicolumn{1}{c|}{-} & - & - & - \\

     MURRE      & - & 87.6 & \multicolumn{1}{c|}{80.1} & - &  21.8 & - & \multicolumn{1}{l}{-} & 94.3 & \multicolumn{1}{c|}{93.5} & - & 64.4 & - \\

    \bf {\sc REaR}      & 36.1 & 93.9 & \multicolumn{1}{c|}{87.3} & \textbf{22.1} & \underline{27.9} & \textbf{34.0} & \multicolumn{1}{l}{\underline{29.6}} & \textbf{98.5} & \multicolumn{1}{c|}{\textbf{97.7}} & \textbf{58.3} & \textbf{73.2} & \underline{74.3} \\
    ~~--Desc.   & 34.7 & 91.1 & \multicolumn{1}{c|}{81.1} & \underline{21.57} & 26.92 & \underline{32.37} & \multicolumn{1}{l}{29.1} & 96.8 & \multicolumn{1}{c|}{95.6} & 51.1 & 55.5 & 66.7 \\
    \hline
\end{tabular}
\label{tab:llm}
\vspace{-1.5em}
\end{table*}

\paragraph{To what extent do gains in retrieval performance from the {\sc REaR} pipeline lead to improvements in downstream end-to-end SQL execution accuracy?}

Ultimately, an advanced retriever is only as valuable as its ability to produce more accurate SQL queries. Table \ref{tab:standard-ete} reveals a direct and positive propagation of retrieval gains to end-to-end performance. Using the UAE dense retriever and Gemini 2.0 Flash on the BIRD dataset, the baseline \textit{``Retrieval (Top 5)''} yields an execution accuracy of 37.87\%. The introduction of the \textbf{Expansion} stage elevates this accuracy to 42.24\% (+4.37 points), indicating that a more complete schema context directly mitigates downstream generation errors. The full {\sc REaR} pipeline with \textbf{Refinement} further improves accuracy to 43.09\%, demonstrating that by increasing the signal-to-noise ratio in the retrieved table set, our pruning mechanism provides a cleaner, higher-fidelity context to the generator, culminating in the best end-to-end performance.


\vspace{-0.25em}
\paragraph{Does the improved retrieval performance of the {\sc REaR} pipeline hold consistently across different SQL generation models?}

To validate the model-agnostic nature of our retrieval framework, we evaluated its impact across three distinct LLMs. The results in Table \ref{tab:standard-ete} demonstrate that the benefits of {\sc REaR} are robust and generalize consistently. For the dense retriever on BIRD, the full \textit{``+Refining''} pipeline improved execution accuracy over the \textit{``Retrieval (Top 5)''} baseline for all models: from 37.87\% to 43.09\% for Gemini 2.0 Flash, from 9.32\% to 11.55\% for Llama-3.2-3B, and from 17.6\% to 17.97\% for Gemma-3-4B. This consistent performance uplift, irrespective of the generator's scale or architecture, confirms that providing a high-quality, well-pruned set of tables is a fundamental improvement at the retrieval level that robustly enhances end-to-end SQL performance.

\vspace{-0.25em}
\paragraph{How much gap does our approach close w.r.t. to human retrieval and human pruning?}
We compare {\sc REaR} against two oracle baselines: Oracle Retrieval, where a human perfectly selects the relevant set of tables, and Oracle Prune, where a human perfectly prunes a candidate set, either from standard retrieval or after our method expansion. Across all three models (Gemini 2.0 Flash, LLaMA 3.2-3B, and Gemma 3 4B~\cite{team2025gemma}) and datasets (BIRD, SPIDER, and MMQA), Oracle Retrieval consistently yields significantly higher accuracy than standard retrievers, highlighting the large gap between current retrieval quality and the upper bound.  This oracle-prune benefit is especially pronounced for weaker models: LLaMA-3.2-3B improves from 9.32\% to 14.69\% (+5.14 points) on BIRD, confirming that retrieval noise is a critical bottleneck for smaller models that lack the implicit robustness of larger LLMs.


Our {\sc REaR} pipeline closes much of this gap by combining recall-oriented expansion with refinement, which effectively prunes irrelevant candidates and brings performance close to the Oracle Prune upper bound. On MMQA with Gemini, Hybrid accuracy rises from 32.96\% (Top-5) to 38.15\% after refinement, compared to 47.08\% for Oracle Prune after expansion recovering over 5.19\% of the potential gain. On SPIDER, refinement achieves 70.9\%, just 4 points below the 74.95\% Oracle score. Even in smaller models, refinement consistently improves over naive expansion and captures much of the oracle-level benefit. These results demonstrate that our method maintains precision while improving recall for multi-table reasoning.


\paragraph{How does the {\sc REaR} pipeline compare to state-of-the-art LLM-based retrieval methods in terms of both retrieval quality and end-to-end execution accuracy?}

A primary objective of this work is to achieve comparable performance with computationally expensive LLM-based retrieval systems. The benchmarks in Table \ref{tab:llm} confirm {\sc REaR}'s success in this regard. On the BIRD dataset, our method records a SQL execution accuracy of 34.00\% with the Gemini model, outperforming the ARM baseline of 30.42\%. On Llama, {\sc REaR} maintains competitive performance at 22.1\%, outperforming ARM (20.6\%) while ReAct degrades severely to just 4.7\%. On retrieval quality, our method achieves superior recall of 93.9\% and full recall of 87.3\%, compared to MURRE's 87.6\% and 80.1\%, respectively. Similarly, on the Spider dataset, our Gemini-based result of 74.3\% is highly competitive with JAR's reported accuracies. The retrieval quality metrics further validate our approach: {\sc REaR} achieves consistently high full recall@k of 87.3\% (BIRD) and 97.7\% (Spider), where missing key tables leads to cascading SQL generation failures. These outcomes are significant, proving that an efficient, non-LLM retrieval architecture can match or exceed the performance of methods that rely on costly iterative LLM reasoning for retrieval.

\vspace{-0.25em}
\paragraph{What computational efficiency advantages does the {\sc REaR} pipeline offer over LLM-based retrieval approaches?}

Beyond raw performance, computational overhead is a critical vector for evaluating real-world viability. As quantified in Table \ref{tab:cost-analysis}, the efficiency gains of our LLM-free retrieval architecture are an order of magnitude. On the BIRD dataset, the {\sc REaR} pipeline requires an average of only 1,533.89 tokens per query for the final SQL generation step. In stark contrast, the LLM-guided ARM retriever consumes 19,747.58 tokens. This represents a \textbf{92.24\% reduction in token consumption}, underscoring the profound cost and latency advantages inherent to our design. This level of efficiency makes {\sc REaR} a far more scalable and economically viable solution for latency-sensitive, production-scale environments.

\begin{table}[!htbp]
\vspace{-0.5em}
\small
\centering
\caption{Cost comparison with LLM-based methods based on number of tokens}
\vspace{-0.5em}
\label{tab:cost-analysis}
\begin{tabular}{l|ccc}
\hline 
\bf Tokens &\bf \#Input&\bf \# Output &\bf \# Total \\
\hline
ARM      &   19307.34   &    440.24          &   19747.58   \\
MURRE      &     16055      &     1600        &  17655  \\
JAR      &       2140.04       &      81.74        &  2221.78  \\
\bf {\sc REaR}     &   1492.47   &  41.41    &   \textbf{1533.89}   \\
\hline
\end{tabular}%
\vspace{-1.5em}
\end{table}

\section{Ablation Studies}

\begin{figure}[!h]
\vspace{-0.5em}
    \centering
    \includegraphics[width=1\linewidth]{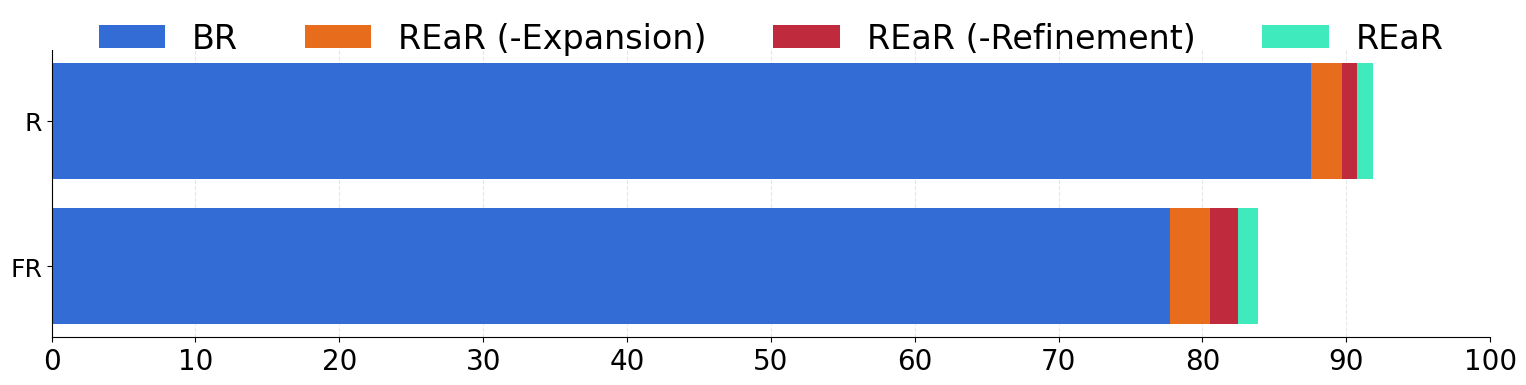}
    \caption{The average recall and Full Recall of base
retrieval (UAE) and our method
with modules removed: {\sc REaR}(-Expansion),
{\sc REaR}(-Refinement), and {\sc REaR} for the top-5 retrieved objects across
Bird, MMQA, and Spider}
    \label{fig:ablations}
    \vspace{-1.5em}
\end{figure}

We conduct ablation studies to isolate the contribution of each component in {\sc REaR}: base retrieval, expansion through joinability discovery, and refinement via attention-inspired pruning. Figure~\ref{fig:ablations} presents the cumulative performance gains, averaged across three datasets with UAE retrieval at each stage.

\paragraph{Expansion.} 
We isolate the effect of expansion by comparing base retrieval (BR) against retrieval with expansion only ({\sc REaR} w/o Refinement). 
Here we naively prune the tables after expansion using query-table similarity computed with the cross encoder. 
Expansion yields \textbf{+4.76 points in PR} (77.72\% $\rightarrow$ 82.48\%) and \textbf{+3.21 points in R} (87.55\% $\rightarrow$ 90.76\%). This confirms joinability-based expansion recovers relevant tables missed by semantic retrieval alone.


\paragraph{Refinement.} 
We isolate the effect of refinement by comparing base retrieval (BR) against retrieval with refinement only ({\sc REaR} w/o Expansion). Refinement yields \textbf{+2.82 points in PR} (77.72\% $\rightarrow$ 80.54\%) and \textbf{+2.14 points in R} (87.55\% $\rightarrow$ 89.69\%). Without expansion, refinement still improves performance by applying cross-encoder scoring to better rank the initial candidate set, filtering out weakly related tables while retaining those most likely to contribute to correct SQL generation.

\paragraph{Combined Effect.} 
The full pipeline ({\sc REaR}) achieves \textbf{83.86\% PR and 91.83\% R}, representing total gains of \textbf{+6.14 points in PR and +4.28 points in R} over the baseline. Comparing {\sc REaR} against each ablated variant reveals the synergistic effect: adding refinement to expansion contributes an additional \textbf{+1.38 points in PR} (82.48\% $\rightarrow$ 83.86\%) and \textbf{+1.07 points in R} (90.76\% $\rightarrow$ 91.83\%), while adding expansion to refinement contributes \textbf{+3.32 points in PR} (80.54\% $\rightarrow$ 83.86\%) and \textbf{+2.14 points in R} (89.69\% $\rightarrow$ 91.83\%).

This dual benefit validates our design: expansion maximizes coverage by discovering joinable tables, while refinement maintains precision by pruning noise through multi-faceted relevance scoring. This structured approach is particularly valuable for smaller language models, which lack the implicit filtering capabilities of larger models and thus benefit more from explicit noise reduction.

\section{Conclusion}
We introduce \textbf{{\sc REaR}}—a three-stage (Retrieve, Expand, Refine) framework that jointly optimizes query–table relevance and table–table joinability. {\sc REaR} retrieves semantically relevant tables, expands with structurally joinable candidates via precomputed column embeddings, and refines with precision-focused pruning, enabling efficient, high-fidelity multi-table retrieval without online LLM calls. Across BIRD, MMQA, and Spider, {\sc REaR} consistently improves retrieval and the resulting SQL execution accuracy. Despite being LLM-free, it performs on par with state-of-the-art LLM-based systems (ARM, JAR) while using fewer tokens, making it practical at scale. These results highlights that explicitly modeling table–table compatibility is crucial for effective multi-table retrieval.

We will extend {\sc REaR} in four directions: (1) richer join patterns, supporting n-ary and conditional joins with join-path discovery under structural constraints and cost-based optimization; (2) learned joinability signals that blend schema metadata, FK structure, and column-embedding similarity to predict valid joins while suppressing spurious ones; (3) adaptive expansion that balances recall and latency by adjusting depth to query complexity via query-aware heuristics or reinforcement learning; and (4) execution-in-the-loop refinement, using lightweight SQL feedback and counterfactual probes to iteratively prune candidates and improve end-to-end SQL accuracy.

\section*{Limitation}
Our work focuses on static relational databases with explicit schema structures, limiting exploration of other data paradigms such as NoSQL databases, knowledge graphs, and dynamically evolving schemas that require real-time index updates. While {\sc REaR} demonstrates effectiveness on merged database corpora, true cross-database retrieval scenarios involving distributed systems with heterogeneous schema conventions remain unexplored. Additionally, our evaluation is confined to single-turn text-to-SQL queries; multi-turn conversational scenarios where context accumulates across interactions and previous retrieval decisions influence subsequent queries present unique challenges our current framework does not address. Extension to multilingual scenarios, databases with implicit or complex join patterns (multi-hop joins, self-joins, non-equi joins), and semi-structured or hierarchical data representations requires further investigation. These limitations suggest important directions for future work in building more robust and generalizable multi-table retrieval systems.
\section*{Ethics Statement}
All datasets used: BIRD, MMQA, and Spider, are publicly available and released under open research licenses. No private, personal, or demographic information is used or inferred in any form. Our pipeline operates solely over structured, non-sensitive tabular data, ensuring privacy and security compliance.\\
We acknowledge that retrieval-based methods can, in theory, be misused to aggregate or generate misleading information. However, our approach is designed exclusively for academic research in question answering and is not intended for deployment in decision-making or surveillance systems.\\
To minimize environmental impact, we reuse pretrained retrievers and perform experiments on limited-scale infrastructure. While these datasets are domain-neutral, inherent topical biases may exist, which we mitigate through standardized evaluation protocols and transparent reporting of all results.\\
This research primarily benefits the NLP community by improving efficiency and interpretability in multi-table reasoning, with minimal risk to end-users. Any future public release of models or code will follow responsible open-source practices, including detailed documentation of intended use to reduce misuse potential. We used AI assistance to help in the writing process.


\bibliographystyle{acl_natbib}
\bibliography{acl_latex}

\appendix
\section{Appendix}
\label{sec:appendix}

\subsection{Query-table relevance}
\label{app:qt relevance}
We employ multiple retrieval strategies to obtain the base tables in order to capture the different aspects of table-query relevance:\\
\textbf{Dense Retrieval.} Dense retrievers encode both the query and table representations into a shared semantic embedding space. Given a query $q$ and a table $t_i$, we compute:

{ \small
\begin{equation*}
\text{score}_{\text{dense}}(q, t_i) = \text{sim}(\mathbf{e}_q, \mathbf{e}_{t_i})
\end{equation*}
}

where $\mathbf{e}_q$ and $\mathbf{e}_{t_i}$ are the embeddings of the query and table respectively, and $\text{sim}(\cdot, \cdot)$ is typically cosine similarity. Dense retrievers excel at capturing semantic similarity and can match queries to tables even when there is no lexical overlap. They are particularly effective at understanding paraphrases, synonyms, and conceptual relationships between the query and table content.

\textbf{Sparse Retrieval.} Sparse retrieval methods score tables based on term frequency and inverse document frequency, capturing exact keyword matches. For a query $q$ and table $t_i$, the sparse retrieval score is computed based on lexical matching:

{ 
\small
\begin{equation*}
\small
\begin{aligned}
\text{score}_{\text{sparse}}(q, t_i) &= \sum_{w \in q} \text{IDF}(w) \cdot \\
&\quad \frac{f(w, t_i) \cdot (k_1 + 1)}{f(w, t_i) + k_1 \cdot (1 - b + b \cdot \frac{|t_i|}{\text{avgdl}})}
\end{aligned}
\end{equation*}
}

where $f(w, t_i)$ is the frequency of term $w$ in table $t_i$, $|t_i|$ is the table length, $\text{avgdl}$ is the average table length, and $k_1$ and $b$ are tuning parameters. Sparse methods provide strong baselines and excel when queries contain specific technical terms or entity names that appear verbatim in table schemas or data.

\textbf{Hybrid Retrieval.} We also explore hybrid approaches that combine dense and sparse signals to leverage both semantic understanding and exact matching capabilities. The hybrid score is computed as a weighted combination:
\begin{equation*}
\small
\begin{aligned}
\text{score}_{\text{hybrid}}(q, t_i) = \alpha \cdot \text{score}_{\text{dense}}(q, t_i) + \\ (1-\alpha) \cdot \text{score}_{\text{sparse}}(q, t_i)
\end{aligned}
\end{equation*}
where $\alpha$ is a hyperparameter balancing the two retrieval paradigms. Hybrid methods aim to capture the complementary strengths of both approaches: the semantic understanding of dense retrievers and the precision of sparse keyword matching.

Standard retrievers compare the query $q$ against individual tables in isolation, selecting the top-$k$ based solely on query-table semantic similarity. It does not consider the implicit relationships between the tables.

\subsection{Prompts and Experimental Setup}
\subsubsection{Experimental Setup}
 We run all our experiments on Nvidia A6000 GPU with 48GB of VRAM, and 120 GB RAM. 
\subsubsection*{Table Description Generation}
We use the following prompt for generating table descriptions before running our pipeline:
\begin{lstlisting}[language=python, basicstyle=\ttfamily\footnotesize, breaklines=true, frame=none, backgroundcolor=\color{gray!10}]
prompt = f"""You are a database documentation expert. Analyze this database table variant and create a comprehensive description for retrieval and search purposes.

Table Name: {table_sample['table_name']}
Variant ID: {table_sample['variant_id']} (Variant #{table_sample['variant_index'] + 1})
Columns: {', '.join(table_sample['columns'])}
Total Rows in this variant: {table_sample['total_rows_in_variant']}

Sample Data (first 10 rows):
{json.dumps(table_sample['sample_data'], indent=2)}

Create a detailed table description that includes:
    1. What this table represents and its main purpose
    2. Columns and their roles


The description should be 2-4 sentences long, informative, and optimized for database retrieval systems. Focus on being descriptive yet concise, and highlight what makes this variant unique.

Table Variant Description:"""
\end{lstlisting}

We use \textsc{Gemini-1.5-Flash} for generating prompts with temperature=0.3 and max tokens=800.

\subsubsection*{SQL Generation}
We use the following prompt for SQL generation:
\begin{lstlisting}[language=python, basicstyle=\ttfamily\footnotesize, breaklines=true, frame=none, backgroundcolor=\color{gray!10}]
prompt = f"""You are an expert SQL query generator. Given the following tables and a natural language question, generate a precise SQL query that answers the question.

AVAILABLE TABLES:
{tables_info}

QUESTION: {question}

INSTRUCTIONS:
1. Analyze the question carefully to understand what information is being requested
2. Identify which tables and columns are needed from the available tables
3. Generate a syntactically correct SQL query that answers the question
4. Use appropriate JOINs if multiple tables are needed
5. Apply proper filtering, grouping, and ordering as required
6. Use the exact column names as shown in the schema
7. Be careful with column names that contain spaces - use backticks or quotes as needed
8. Return ONLY the SQL query without any explanation or markdown formatting

SQL QUERY:"""
\end{lstlisting}

For all LLMs, we use temperature 0.2, top-k sampling with $k=1$, top-p sampling with $p=1.0$, maximum token length of 500, and random seed 42. We access GPT-4o-mini via OpenAI's API, Gemini-2.0-Flash via Google's API~\cite{team2023gemini}, and other models via DeepInfra's API.
\subsection{Joinability analysis}

We adapt the column-level joinability approach from DeepJoin \cite{dong2023deepjoin} to discover additional relevant tables. While DeepJoin fine-tunes a language model on the table corpus, we use a general-purpose embedding model to avoid the computational overhead of fine-tuning while achieving comparable performance. We adopt DeepJoin's column description format:

\begin{table}[h]
\small
\begin{tabular}{l|l}
\hline
\textbf{Name} & \textbf{Pattern} \\
\hline
title-colname-stat-col & \$table\_title\$.\$colname-stat-col\$ \\
\hline
\end{tabular}
\end{table}

We construct a column-level vector index by embedding all columns in the database and indexing them using FAISS for efficient approximate nearest neighbor search. Given the initially retrieved tables, we query this index to find columns with high similarity to those in the retrieved set, explicitly excluding tables already present to avoid redundancy. This yields a ranked list of candidate tables based on column-level joinability.

To select tables for expansion, we rerank these candidates using Jina-Reranker-v2 based on query-table relevance and add the top-3 tables to the retrieved corpus. We choose to evaluate on 3 of the top models in the MTEB leaderboard: Qwen3-embedding-0.6B \cite{zhang2025qwen3}, gemini-embedding-001 \cite{lee2025gemini}, and bge-large-en. Table~\ref{tab:expansion-methods} compares embedding models for this task on MMQA with e5-mistral retrieval.
\begin{table}[h]
\small
\centering
\caption{Comparison of various expansion models evaluated on MMQA using e5-mistral. }
\label{tab:expansion-methods}
\begin{tabular}{l|lll}\hline
\bf Model & \bf Precision &\bf Recall &\bf Full\\ 
\bf  & \bf &\bf  &\bf Recall\\ \hline
bge-large-en  & 24.31 &  87.39 &   74.73     \\
gemini-embedding-001      & 24.47  & 87.85  & 75.83  \\
Qwen3-embedding-0.6B     &  23.91 & 85.78  & 72.56 \\
 \hline
\end{tabular}
\end{table}
We select bge-large-en for this component due to its strong balance of performance and efficiency. While gemini-embedding-001 achieves marginally higher recall (+0.46pp) and full recall (+1.1pp), bge-large-en offers comparable performance with significantly lower computational cost and greater accessibility for reproduction.

\subsection{Pruning Techniques}

We evaluated multiple pruning strategies to optimize table selection, ultimately adopting the Refinement approach described in \cref{sec:pruningab}. Table~\ref{tab:pruning-methods} presents results for various pruning methods on MMQA using the UAE retriever. We report on MMQA rather than BIRD or Spider because the performance differences between methods are more pronounced on this dataset, allowing clearer comparison of pruning strategies. The experimental variants are described below:

\subsubsection{Pruning With Alpha-Beta}
\label{sec:pruningab}
In this method we compute the query-table and table-table scores using a cross-encoder model (Jina reranker). Based on the intuition that columns across two tables are joinable will have high similarity, we compute column-column scores pair-wise between all the tables across the retrieved table corpus and take the maximum score across the columns of a pair of tables as the final score between those two tables. We also compute the table-query score using the same cross-encoder and finally get a final score as follows:

{
\vspace{-1.0em}
\small
\begin{equation*}
S(T_i) = \alpha.C(Q, T_i) + \frac{1}{n}\beta.\sum_{j\in N} C(T_j, T_i).C(T_j, Q)
\vspace{-0.5em}
\end{equation*}
}

where $S(T_i$ is the score of the $i^{th}$ table in the retrieved corpus, $\alpha, \beta$ are the weighing values, $C(.,.)$ is the cross encoder score, $N$ is the neighborhood of table i and $n$ are the total tables in the neighborhood. We stuck with the values $\alpha = 0.6$ and $\beta = 0.4$ after some hyperparameter tuning.

\begin{table}[h]
\small
\centering
\caption{\small Comparison of various pruning methods evaluated on MMQA using UAE. }
\label{tab:pruning-methods}
\begin{tabular}{l|ccc}\hline
\bf Method & \bf Precision &\bf  Recall &\bf Full Recall \\ \hline
Alpha-Beta  & 36.48 &  82.85 &   66.07 \\
Adaptive      & \textbf{44.48}  & 80.48  & 62.11  \\
Max Pruning$^1$     &  36.15 & 82.89  & 66.21  \\
Max Pruning$^2$    &  36.22 & 81.95  & 64.59  \\
Refinement      & 36.86  & \textbf{83.19 } &  \textbf{66.61} \\ \hline
\end{tabular}
\end{table}

\begin{table*}[t]
\centering
\small
\setlength{\aboverulesep}{0pt}
\setlength{\belowrulesep}{0pt}
\setlength{\tabcolsep}{5.5pt}
\caption{Retrieve, Expand, and Refine comparison grouped by number of tables.}
\vspace{-0.5em}
\label{tablewise-results}
\begin{tabular}{@{}c c l  *{3}{cccc} *{3}{cccc}@{}}
\toprule
\bf \#Table &\bf  \#Q & \bf Approach & 
\multicolumn{4}{c}{\bf Dense} & \multicolumn{4}{c}{\bf SPARSE} & \multicolumn{4}{c}{\bf Hybrid} \\
\cmidrule(lr){4-7}\cmidrule(lr){8-11}\cmidrule(l){12-15}
 &  &  & \bf R & \bf P & \bf FR & \bf SQL & \bf R & \bf P & \bf FR & \bf SQL & \bf R & \bf P & \bf FR & \bf SQL \\
\cmidrule(l){4-15}
&&& \multicolumn{12}{@{}c}{\textbf{BIRD}}\\
\cmidrule(l){4-15}
\multirow{3}{*}{\bf 1} & \multirow{3}{*}{\bf 361} &\bf Retrieve & 96.7 & 20.8 & 96.7 & 46.7 & 97.8 & 20.5 & 97.8 & 49.6 & 98.9 & 20.8 & 98.9 & 50.3 \\
&  &\bf Expand  & 97.9 & 13.0 & 97.9 & 48.1 & 99.5 & 12.8 & 99.5 & 49.8 & 99.5 & 12.8 & 99.5 & 49.9 \\
& &\bf Refine   & 97.8 & 19.6 & 97.8 & 50.0 & 99.2 & 19.8 & 99.2 & 49.8 & 99.2 & 19.8 & 99.2 & 54.3 \\
\cmidrule(l){4-15}
\multirow{3}{*}{\bf 2} & \multirow{3}{*}{\bf 924} &\bf  Retrieve & 91.8 & 38.7 & 85.5 & 37.8 & 91.6 & 37.8 & 84.9 & 35.7 & 92.8 & 38.3 & 86.2 & 37.9 \\
 & &\bf  Expand   & 97.6 & 26.6 & 95.6 & 40.8 & 98.4 & 25.0 & 97.0 & 39.5 & 97.9 & 24.9 & 96.2 & 39.6 \\
 & &\bf  Refine   & 95.1 & 38.0 & 90.7 & 39.1 & 94.9 & 38.0 & 90.0 & 40.5 & 94.6 & 37.8 & 89.5 & 39.6 \\
\cmidrule(l){4-15}
\multirow{3}{*}{\bf 3+} & \multirow{3}{*}{\bf 249} &\bf Retrieve  & 77.7 & 50.1 & 52.6 & 25.1 & 78.9 & 50.0 & 47.0 & 23.7 & 82.0 & 52.3 & 54.4 & 26.2 \\
 & &\bf Expand    & 92.1 & 37.1 & 78.2 & 39.1 & 92.5 & 36.5 & 78.7 & 34.9 & 91.6 & 36.2 & 76.7 & 31.7 \\
 & &\bf Refine    & 84.3 & 52.9 & 59.4 & 32.2 & 84.0 & 52.6 & 59.0 & 27.7 & 83.6 & 52.4 & 57.8 & 25.7 \\
\cmidrule(l){4-15}
&&&\multicolumn{12}{@{}c}{\textbf{MMQA}}\\
\cmidrule(l){4-15}
\multirow{3}{*}{\bf 2} & \multirow{3}{*}{\bf 2591} & \bf Retrieve & 76.2 & 30.5 & 58.3 & 34.65 & 77.7 & 31.1 & 59.4 & 35.39 & 80.9 & 32.4 & 64.4 & 35.9 \\
  & &\bf  Expand  & 89.1 & 22.3 & 79.5 & 43.4 & 89.9 & 22.5 & 80.7 & 42.18 & 90.4 & 22.6 & 81.6 & 39.9 \\
 & & \bf Refine  & 85.1 & 34.0 & 72.1 & 42.2 & 85.6 & 34.2 & 72.6 & 41.8 & 84.0 & 33.6 & 69.9 &  41.3 \\
\cmidrule(l){4-15}
\multirow{3}{*}{\bf 3+} & \multirow{3}{*}{\bf 721} & \bf Retrieve & 65.7 & 40.3 & 30.5 & 25.17 & 69.5 & 42.6 & 34.1 & 23.71 & 69.4 & 42.6 & 31.1 & 24.1 \\
 & & \bf Expand & 83.7 & 32.2 & 61.9 & 31.8 & 85.8 & 33.0 & 66.9 & 32.78 & 84.7 & 32.7 & 63.7 & 33.5 \\
 & & \bf Refine & 76.4 & 47.0 & 47.0 & 31.1 & 77.6 & 47.7 & 50.5 & 32.2 & 72.7 & 44.8 & 38.7 & 31.8 \\
\cmidrule(l){4-15}
&&&\multicolumn{12}{@{}c}{\textbf{SPIDER}}\\
\cmidrule(l){4-15}
\multirow{3}{*}{\bf 1} & \multirow{3}{*}{\bf 585} & \bf Retrieve & 96.8 & 19.8 & 96.8 & 75.5 & 81.2 & 16.2 & 81.2 & 66.9 & 87.7 & 17.5 & 87.7 & 52 \\
 & & \bf Expand   & 99.2 & 12.6 & 99.2 & 76.0 & 99.2 & 12.4 & 99.2 & 78.0 & 96.1 & 12.0 & 96.1 & 78.6 \\
 & & \bf Refine & 98.6 & 20.2 & 98.6 & 76.0 & 96.4 & 19.3 & 96.4 & 77.4 & 95.9 & 19.2 & 95.9 & 77.7 \\
\cmidrule(l){4-15}
\multirow{3}{*}{\bf 2} & \multirow{3}{*}{\bf 383}  & \bf Retrieve &  95.6 & 38.5 & 91.1 & 66.3 & 82.1 & 32.9 & 64.2 & 48.3 & 80.3 & 32.1 & 61.4 & 39.9  \\
  & & \bf Expand  & 99.5 & 25.0 & 99.0 & 65.3 & 99.1 & 24.8 & 98.2 & 66.1 & 97.7 & 24.4 & 95.3 & 65.3 \\
 & & \bf Refine & 97.6 & 39.3 & 95.3 & 65.3 & 98.4 & 39.4 & 96.9 & 66.8 & 97.5 & 39.0 & 95.0 & 65.8 \\
\cmidrule(l){4-15}
\multirow{3}{*}{\bf 3+} & \multirow{3}{*}{\bf 66}  & \bf Retrieve & 93.9 & 58.6 & 83.3 & 34.9 & 75.0 & 46.4 & 31.8 & 30.3 & 80.1 & 49.4 & 42.4 & 33.3 \\
  & & \bf Expand  & 100.0 & 38.9 & 100.0 & 36.4 & 100.0 & 38.7 & 100.0 & 36.4 & 97.5 & 37.8 & 92.4 & 39.4 \\
 & & \bf Refine  & 95.5 & 59.6 & 86.4 & 36.5 & 97.5 & 60.3 & 92.4 & 37.9 & 95.5 & 59.1 & 86.4 & 40.9 \\
\bottomrule
\end{tabular}
\end{table*}

\subsubsection{Adaptive Pruning}
We employ the same scoring mechanism described in \cref{sec:pruningab}, but with an adaptive thresholding strategy. Rather than selecting a fixed top-k tables, we compute the mean score across all candidates and retain only those tables scoring above this threshold. This dynamic selection adjusts the retrieved set size based on score distribution: high-quality candidates yield smaller, focused sets, while ambiguous queries produce larger sets.

This approach trades recall for precision. By filtering out below-average candidates, we reduce noise in the final corpus at the potential cost of excluding marginally relevant tables. In practice, this adaptive strategy is particularly effective when precision is critical for downstream SQL generation.

\subsubsection{Max Pruning}
We compute table-table and query-table scores as described in \cref{sec:pruningab}, but aggregate neighborhood scores using maximum instead of mean. The intuition is that a table's relevance should be determined by its strongest connection: if it is highly similar to even one neighbor that is also query-relevant, it likely contains useful information for SQL generation.

We evaluate two variants of this max-pooling strategy, {\small $S(T_i) =$}

{
\vspace{-1.0em}
\small
\begin{equation*}
\begin{aligned}
\alpha \cdot C(Q, T_i) + \beta \cdot \max_{T_j \in \mathcal{N}(T_i)} \max(C(T_j, T_i),C(T_j, Q))
\end{aligned}
\vspace{-0.5em}
\end{equation*}
\vspace{-0.5em}
\begin{equation*}
\begin{aligned}
\max\left(C(Q, T_i), \max_{T_j \in \mathcal{N}(T_i)} \max(C(T_j, T_i),C(T_j, Q))\right)
\end{aligned}
\end{equation*}

}
where $S(T_i)$ is the score of table $T_i$, $C(\cdot,\cdot)$ denotes cross-encoder similarity, $\mathcal{N}(T_i)$ is the neighborhood of $T_i$, and $\alpha, \beta$ are weighting hyperparameters. The first variant combines query relevance and neighborhood strength linearly, while the second uses pure max-pooling across all relevance signals.

\subsection{Table Wise Multitable Retrieval}

Table \ref{tablewise-results} presents a granular analysis of {\sc REaR}'s performance stratified by query complexity, measured by the number of gold tables required for correct SQL generation.

The results reveal a clear correlation between query complexity and the magnitude of improvement delivered by our pipeline. For single-table queries, where retrieval is relatively trivial, the baseline \textbf{"Retrieve"} approach already achieves near-perfect recall (96.7-98.9\% across datasets), with expansion and refinement providing marginal gains primarily in SQL execution accuracy. The benefits of {\sc REaR} become more pronounced for two-table queries, where the expansion stage consistently boosts full recall by 8-13 percentage points across all datasets (e.g., BIRD Dense for 2 tables: 85.5\% → 95.6\%), while refinement successfully recovers precision losses without sacrificing these recall gains. 

Most notably, for the most challenging 3+ table queries, expansion delivers substantial improvements. For example, on BIRD, full recall increases from 52.6\% to 78.2\% for dense retrievers, demonstrating that join-aware expansion is critical for complex multi-table reasoning. The refinement stage demonstrates its effectiveness in restoring precision while preserving the recall gains achieved through expansion. 

Across all query complexities, refinement consistently improves precision by 10-15 percentage points, on BIRD with 2-table queries, precision increases from 26.6\% to 38.0\% for dense retrievers, while full recall drops only marginally from 95.6\% to 90.7\%. This precision-recall trade-off is particularly valuable for 3+ table queries, where refinement boosts precision from 37.1\% to 52.9\% on BIRD, indicating successful removal of weakly related tables introduced during expansion.

\end{document}